\documentclass[twocolumn,showpacs,pre,amsmath,amssymb]{revtex4}

\usepackage{graphicx}
\usepackage{dcolumn}
\usepackage{bm}

\begin{document}

\title{
First order isotropic - smectic-A transition in liquid crystal-aerosil gels
} 

\author{M.K.~Ramazanoglu}
\affiliation{Department of Physics, University of Toronto, Toronto, Ontario M5S 1A7, Canada}

\author{P.S.~Clegg}
\affiliation{Department of Physics, University of Toronto, Toronto, Ontario M5S 1A7, Canada}

\author{R.J.~Birgeneau}
\affiliation{Department of Physics, University of Toronto, Toronto, Ontario M5S 1A7, Canada}

\author{C.W.~Garland}
\affiliation{Department of Chemistry, Massachusetts Institute of Technology, Cambridge, MA 02139}

\author{M.E.~Neubert}
\affiliation{Liquid Crystal Institute, Kent State University, Kent,
Ohio, 44242}

\author{J.M.~Kim}
\affiliation{Liquid Crystal Institute, Kent State University, Kent,
Ohio, 44242}

\date{\today}

\begin{abstract}
The short-range order which remains when the isotropic to smectic-A transition is perturbed by a
gel of silica nanoparticles (aerosils) has been studied using high-resolution synchrotron x-ray
diffraction. The gels have been created \textit{in situ} in decylcyanobiphenyl (10CB), which has a
strongly first-order isotropic to smectic-A transition. The effects are determined by detailed
analysis of the temperature and gel density dependence of the smectic structure factor. In
previous studies of the continuous nematic to smectic-A transition in a variety of thermotropic
liquid crystals the aerosil gel appeared to pin, at random, the phase of the smectic density
modulation. For the isotropic to smectic-A transition the same gel perturbation yields different results.
The smectic correlation length decreases more slowly with increasing random field variance in good
quantitative agreement with the effect of a random pinning field at a transition from a uniform
phase directly to a phase with one-dimensional translational order. We thus compare the
influence of random fields on a \textit{freezing} transition with and without an intervening
orientationally ordered phase.
\end{abstract}

\pacs{64.70.Md,61.30.Eb,61.10.-i}

\maketitle

\section{Introduction}
\label{sec:intro}

The formation of the smectic-A (SmA) phase is intriguing because the resulting 
one-dimensional density modulation
lacks true long-range order~\cite{Peierls, L_and_L}. Much
effort has been devoted to experimental characterization and theoretical
description of the SmA phase of thermotropic liquid crystals~\cite{G_and_N, N_and_T}. More
recently, the effects of quenched random disorder on the nematic (N) - SmA transition have been studied. The
interplay between the thermal fluctuations which disrupt the long-range order and static
fluctuations due to quenched randomness is fascinating. Experimentally,
quenched randomness can be introduced via a porous random environment such as aerogel. In order to
probe weaker disorder the rigid aerogel has been replaced by a hydrogen-bonded gel of aerosil
particles. In both cases the N - SmA transition has been studied using
calorimetry~\cite{Aerogel_Cp, Aerosil_Cp}, 
deuterium NMR~\cite{Aerogel_DNMR, Aerosil_DNMR} and
x-ray diffraction~\cite{Aerogel_X, Aerosil_X}. In 
addition, results have been presented for the isotropic (I)
- SmA transition for a liquid crystal confined in an aerogel~\cite{Aerogel_IA}. 
Here we present a comprehensive x-ray diffraction study for an aerosil gel.

For many liquid crystals the SmA phase forms from the orientationally ordered N
phase. There is coupling between the order parameters and fluctuations of the two phases.
As the SmA phase becomes more favorable at higher temperatures, due to changes in the 
molecule or mixture composition, the N-SmA transition occurs
while the N order parameter is less and less saturated. The transition into the SmA phase enhances the N
order. This coupling between order parameters leads to a tricritical point in the phase 
diagram~\cite{dG_and_P}. As the
stability of the SmA phase is further enhanced a triple point occurs in the phase diagram
and the N phase ceases to be stable. Beyond this point, 
orientational order occurs along with positional order in the SmA structure. 
The pretransitional fluctuations of the SmA order in the I phase occur over
a sphere of ordering wave vectors corresponding to the layer spacing. The
transition is thought to be first order due to the contribution of these fluctuations as described by
Brazovskii~\cite{Brazovskii, Anisimov}. The observation that the I-SmA transition is strongly
first order for 10CB~\cite{Marynissen,Thoen_p} may well be due to the de Gennes
coupling between smectic and nematic
order parameters in the isotropic phase~\cite{Gohin}.

The cyanobiphenyls are a well-studied family of thermotropic liquid crystals~\cite{Thoen}. 
These molecules have
a flexible aliphatic tail the length of which
influences the stability of the SmA phase. Both
decylcyanobiphenyl (10CB) and dodecylcyanobiphenyl (12CB) exhibit transitions directly from the
I phase to the SmA phase. Since, in establishing 
the SmA order, both orientational and positional order must be developed, one might expect this
transition 
to be especially strongly perturbed by quenched random disorder.
Thixotropic (hydrogen bonded) gels can be created by dispersing fine, hydrophilic aerosil particles in the 
liquid crystal.
These silica nanoparticles hydrogen bond together creating a flexible structure which can
substantially perturb the interpenetrating liquid crystal material. 
The pore volume fractions range from 0.77 to 0.99.

Previous studies of the N-SmA transition in liquid crystal-aerosil gels have implied that the
main effect of the aerosils is to pin the phase of the SmA density wave~\cite{Aerosil_X, Germano2,
Clegg1}. Specifically, both the size of the SmA
domains and the detailed wave vector dependence of the x-ray structure factor have been in quantitative
agreement
with the predictions of a random pinning field model. In this paper we are reporting the results of a
study of the random field model at
a first-order transition. As first articulated by Imry and Ma, the simplest understanding of the 
effects of random fields comes
from considering a balance between the benefit of aligning with the local random field and the
cost of forming a domain wall~\cite{I_and_M}. This model predicts differing responses for transitions
which break discrete and continuous symmetries. The situation becomes more complicated for
first-order transitions due to the possibility of two-phase coexistence~\cite{I_and_W}. Domain walls
can exist between ordered and disordered regions as well as between ordered domains. 
A guide to the likelihood of two-phase coexistence is whether the disorder causes a
substantial suppression of the phase transition temperature. Should it do so, it is likely
that \textit{large} variations of transition temperatures will occur locally giving 
two-phase coexistence so that the sharp transition will then be smeared out. Two-phase coexistence will
alter the penalty for forming domain walls. Substantial two-phase coexistence results in sharp
boundaries between ordered and disordered regions. This has similarities to a transition which
breaks a discrete symmetry with random fields. One might speculate that 
long-range order could be observed for sufficiently
weak disorder in $d=3$. Without significant two-phase coexistence the domain walls will be broad
and the response should correspond to expectations for a 
transition which breaks a continuous symmetry in the presence of random fields.

In reference~\cite{Aerogel_IA} a silica aerogel is used to impose quenched disorder on 10CB, and
similar previous studies of the liquid crystal
650BC are also discussed. Both of these materials exhibit transitions directly from the I to the SmA phase.
The correlation lengths are observed to increase discontinuously at the transition temperature and
to saturate quickly. Meanwhile, the scattering intensity increases only slowly with decreasing
temperature. 
The transition is observed to remain first order under this strong
confinement and this is confirmed via optical microscopy of the coexisting phases. It is surprising
that the correlation length remains almost constant as the SmA order goes from
the two phase coexistence region to the uniform phase region. The transition temperature is found to remain
roughly unchanged.

Here we show that the liquid crystal-aerosil gel also retains only short-range SmA order deep into
the SmA phase temperature region. In common
with previous studies, a range of gel densities have been prepared and characterized using 
high-resolution x-ray diffraction. The results demonstrate that our quantitative understanding of the
N-SmA transition naturally extends to
this new class of transition. In Sec.~\ref{sec:pre} we outline preliminary details including the
differences for x-ray diffraction between the I-SmA transition and the N-SmA transition and
the experimental techniques employed.
Section~\ref{sec:R_and_A} gives the results and analysis. The quantitative comparison between
these results and existing theoretical models is discussed in Sec.~\ref{sec:dis} and
conclusions are drawn.

\section{Preliminary Details}
\label{sec:pre}

\subsection{X-ray Structure Factor}

The x-ray scattering line shape approaching the N-SmA transition~\cite{Ocko} and deep within 
the SmA phase~\cite{Als-Nielsen} have
been characterized in great detail. In the former case the data for a wide range of liquid
crystalline materials have been found to be described by the structure factor
\begin{equation}
S(\mathbf{q}) = \frac{\sigma_0}{1 + \xi_{\|}^2 (q_{\|} - q_0)^2 + \xi_{\bot}^2 q_{\bot}^2
+ c \xi_{\bot}^4 q_{\bot}^4}
\label{N_to_SmA}
\end{equation}

\noindent where $q_0 = 2\pi / d$ is the wave vector where the SmA peak occurs corresponding to a
layer spacing $d$. The correlation length $\xi_{\|} (\xi_{\bot})$ and wave vector components
$q_{\|} (q_{\bot})$ are parallel (perpendicular) to the nematic director. The susceptibility is
$\sigma_0$, while $c$ ranges between 0.25 and 0 and gives the scale of the fourth-order correction.
The second-order $q_{\bot}$ contribution to the line shape arises due to the influence of nematic
fluctuations on the formation of the SmA state~\cite{Litster}. 
Far below the N-SmA transition these fluctuations
are constrained to be perpendicular to the smectic layers. In the case of finite correlation lengths,
$\xi_{\|}$ and $\xi_{\bot}$, the structure factor then can be written approximately as
\begin{equation}
S(\mathbf{q}) = \frac{\sigma_0}{1 + \xi_{\|}^2 (q_{\|} - q_0)^2
+ \xi_{\bot}^4 q_{\bot}^4}
\label{I_to_SmA}
\end{equation}

\noindent In the case of infinite correlation lengths, the structure factor has been described by
Caill\'{e}~\cite{Caille} and varies with $(q_{\|} - q_0)^{-2 + \eta}$ and $q_{\bot}^{-4 + 2\eta}$ 
respectively ($\eta$ depends on the elastic constants)~\cite{Als-Nielsen}.

Experimental determination of these structure factors for pure liquid crystals 
was carried out in an applied magnetic
field. In the N phase all regions of the sample become aligned allowing detailed characterization
of the pretransitional smectic fluctuations or the algebraic decay within the SmA state.
Equivalent studies are not possible for the I-SmA transition since there is no possibility of
magnetic field alignment in the N state. Hence, only very preliminary studies have been made of
the line shape for 12CB~\cite{C_and_S}. In what follows, we analyze our results assuming 
that the SmA
fluctuations are either similar to those in nonylcyanobiphenyl (9CB) 
as given by~Eq.(\ref{N_to_SmA}) or that 
the scattering is described by Eq.(\ref{I_to_SmA}). At the I-SmA transition the orientational order is
caused by the formation of the SmA state. As a result, we anticipate that the nematic director will
be constrained to lie perpendicular to the smectic layers. By this argument Eq.(\ref{I_to_SmA})
appears to be a good approximation to the structure factor.

\subsection{Experimental Techniques}

The 10CB was 
synthesized at Kent State University. Type-300 aerosil was supplied by Degussa~\cite{Degussa} 
and was dried
prior to use. The amount of aerosil in each sample was quantified by the parameter
$\rho_S$, the mass of aerosil divided by the volume of liquid crystal.
X-ray diffraction
experiments were carried out at the National Synchrotron Light Source at Brookhaven National
Laboratory. The beam lines used were X20A, X20C and X22A. 
Liquid crystal samples containing dispersed
aerosils were in the solid phase following long periods 
at room temperature and so prior to study were remixed with pure ethanol, 
sonicated and then dried for several days on a hot plate.
The gel samples were placed in 5-mm radius disks which have kapton
windows~\cite{Park}. The kapton x-ray diffraction signal has a powder peak outside the range 
of interest for these experiments.
Temperature control is via a PI controller to within $\pm 0.05$ K. Each sample was kept in the I
phase at a temperature around 30 K above the I-SmA transition for 6 hours after being transferred
to the holder. The number of scans for each sample was kept limited to avoid undue x-ray damage.

\begin{figure}
\includegraphics[scale=0.3]{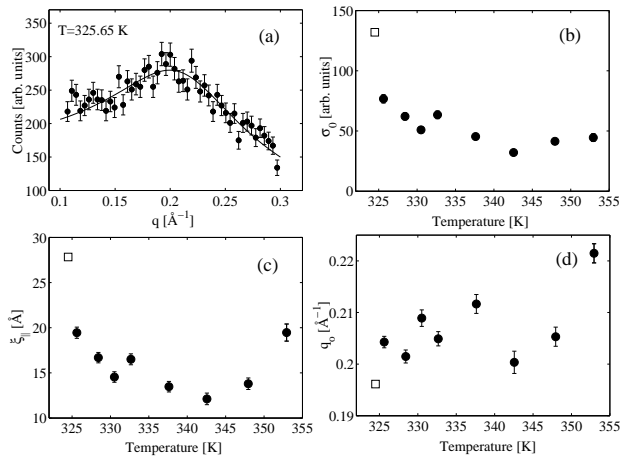}
\caption{\label{Bulk} Measurements and analysis of pretransitional smectic fluctuations in bulk 10CB.
Panel (a) shows a typical scan of the scattering intensity versus wave-vector transfer. The solid line is
the result of a fit with Eq.(\ref{N_to_SmA}). Panels (b, c, d) show susceptibility, parallel 
correlation length, and peak
position resulting from fits to the data. The open symbol is in the region of coexistence of
isotropic and smectic phases.}
\end{figure}

\section{Results and analysis}
\label{sec:R_and_A}

\begin{figure}[b]
\includegraphics[scale=0.5]{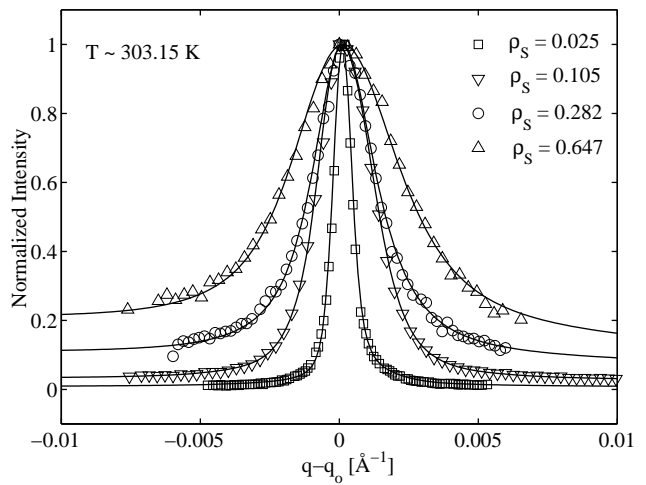}
\caption{\label{scans} Normalized x-ray scattering intensity versus wave-vector transfer for
10CB-aerosil gels due to short-range SmA order at a temperature $\sim 20$ K below the effective
$T^*_{IA}$ value. The lines are the results of fits with a model
described in the text.}
\end{figure}

Figure~\ref{Bulk} presents our measurements and analysis of the pretransitional fluctuations in the 
isotropic phase of 10CB.
Pretransitional smectic
fluctuations are extremely weak in the pure material and were never evident in the gel samples.
Figure~\ref{Bulk}(a) shows a typical scan; the peak is very broad and the solid line is the result
of a fit to Eq.(\ref{N_to_SmA}) varying $\sigma_0$, $q_0$, and $\xi_{\|}$ and using the known 9CB 
characteristics for $\xi_{\bot}(\xi_{\|})$ and $c(\xi_{\|})$. The parameters from such fits
are displayed in panels b,c,d of Fig.~\ref{Bulk}. There is no indication that the susceptibility,
$\sigma_0$, or the parallel
correlation length, $\xi_{\|}$ are about to diverge. 
The weakness of the fluctuations immediately prior to the
transition strongly suggest that this is not a typical fluctuation driven first-order transition.
The observed behavior is likely due to the de Gennes coupling between nematic and smectic order
parameters.

The 10CB-aerosil sample compositions studied are listed in Table~\ref{tab:Tna}.
Measurements were made from a temperature ($T \sim 350$ K) well above the I-SmA transition 
($T_{IA} = 324.5$ K). However, the SmA line shape for 10CB-aerosil samples
only appeared at a pseudotransition temperature with a substantial correlation length already
established. Typical scans are presented in Fig.~\ref{scans}. The peaks are all
broader than the instrumental resolution. The width increases with increasing gel density but
varies little with temperature. Data were taken down to $T = 298$ K, which is well below the
bulk material crystallization temperature~\cite{BDH} ($T_{K} = 317.1$ K).
 The lines in Fig.~\ref{scans} corresponds to a model that
will be described below. Two samples, $\rho_S = 0.025$ and $0.282$, were measured on beam-line 
X22A where
slightly higher instrumental resolution could be achieved. Typically, the peak position in the SmA phase was 
$q_0 = 0.1776$ \AA$^{-1}$, rising just below the transition and dropping slightly at lower temperatures.

For previous x-ray studies of the N-SmA transition in aerosil gels, the line shape was taken to be
composed of two contributions~\cite{Aerosil_X, Clegg1, Park, Clegg2}. 
The first part is the thermal term, Eq.(\ref{N_to_SmA}), which is identical to that of
the pure material, while the second part is the random-field term, which has the form of the thermal
term squared. The total structure factor is
\begin{eqnarray}
\label{two_terms}
S(\mathbf{q}) = & \sigma_1 S^{T}(\mathbf{q}) + S^{RF}(\mathbf{q})\\
S^{RF}(\mathbf{q}) = & a_2 (\xi_{\|}\xi_{\bot}^2) [S^{T}(\mathbf{q})]^2 \nonumber
\end{eqnarray}

\noindent where $S^{T}(\mathbf{q})$ corresponds to Eq.(\ref{N_to_SmA})
for the N-SmA transition with the numerator set equal to 1
and the denominator of $S^{RF}(\mathbf{q})$ is the square of the denominator of 
$S^{T}(\mathbf{q})$. $\sigma_1$ and $a_2$ are the amplitudes for the thermal and random-field terms
respectively.
Since the samples are comprised of a random arrangement of ordered domains this structure factor
has to be averaged over all orientations. Prior to comparing the model with the data this spherically 
averaged
line shape must be convoluted with the measured instrumental resolution function. Both the integral over
orientations and the convolution were performed numerically using commercially available routines.

\begin{figure}[b]
\includegraphics[scale=0.45]{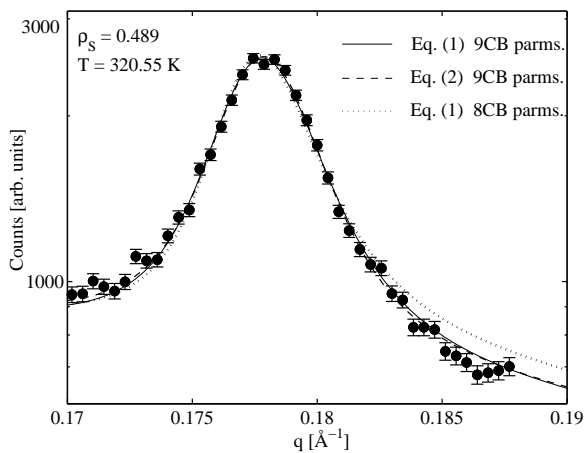}
\caption{\label{line_shape} X-ray scattering intensity due to SmA correlations. The three lines
represent three different models where the assumptions have been changed. The models and the
assumptions are described in the text.}
\end{figure}

\begin{figure}[b]
\includegraphics[scale=0.3]{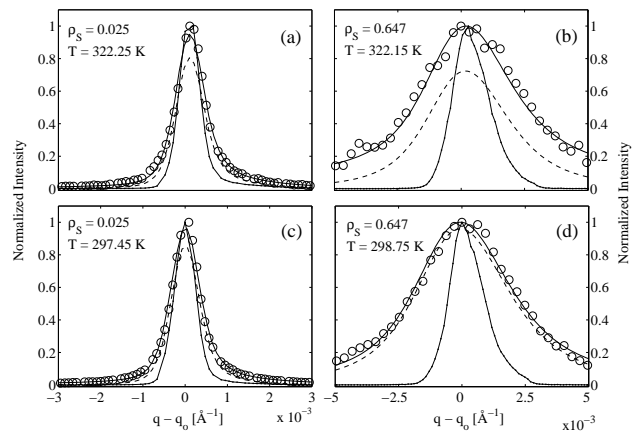}
\caption{\label{4pictures} The variation in the random field contribution (dashed line) to
the x-ray scattering intensity (open circles) with temperature and $\rho_S$. 
The experimental resolution corresponding to each
scan is shown as dots joined by a solid line.}
\end{figure}

Figure~\ref{line_shape} shows the results of fits to a typical profile. The dotted and 
solid lines result from 
applying Eq.(\ref{two_terms}) to the 10CB-aerosil data using the form of
Eq.(\ref{N_to_SmA}) for $S^{T}(\mathbf{q})$. Here, as elsewhere~\cite{Aerosil_X, Park, Clegg1, Clegg2}, 
we take $\xi_{\|}$ as a
parameter to be fitted and we assume that $\xi_{\bot}(\xi_{\|})$ and $c(\xi_{\|})$ remain the 
same as for the pure material~\cite{Leheny_p}. The behavior of $\xi_{\|}$, $\xi_{\bot}$ and $c$ for 
10CB are not
known. The dotted and the solid line in Fig.~\ref{line_shape} assume that 8CB behavior and 9CB 
behavior for $\xi_{\bot}(\xi_{\|})$ and $c(\xi_{\|})$ are
followed respectively~\cite{Ocko_Thesis}. 
It is evident that either fit gives a reasonable description of the data, with
the use of 9CB parameters yielding a somewhat better fit. 
\textbf{This model using the 9CB relationships for
$\xi_{\bot}(\xi_{\|})$ and $c(\xi_{\|})$ has been used to analyze all of
the 10CB-aerosil results.}
The fit parameters are $q_0, \sigma_1, a_2$, and $\xi_{\|}$, whose optimum
values are determined using a least-squares fitting procedure. The dashed line in 
Fig.~\ref{line_shape} shows the result of a fit with Eq.(\ref{two_terms}) to the 10CB-aerosil data 
using the form of
Eq.(\ref{I_to_SmA}) for $S^{T}(\mathbf{q})$. Again $\xi_{\bot}(\xi_{\|})$ is assumed to remain the
same as in bulk 9CB; however in this case the transverse tails of the peak are assumed to have
changed. This model also gives a good correspondence to the data. Use of a model based on Eq.(\ref{I_to_SmA}) 
for the form of $S^{T}(\mathbf{q})$
systematically modifies the fit parameters compared to those presented here. There is a small shift in
favor of the thermal term and the correlation length decreases. The latter feature is returned to in more
detail below.

\begin{table*}
\caption{\label{tab:Tna} Parameters for bulk 10CB and eight gel samples. Shown are the density, 
$\rho_S$, the
mass of aerosil per unit volume of liquid crystal; the pore volume fraction $\Phi$; the low
temperature parallel correlation length $\xi_{\|}$; the low-temperature ratio of the random-field amplitude
$a_2$ to the thermal amplitude $\sigma_1$; the pseudotransition temperature $T_{IA}^*$, and the
exponent $x$ describing the temperature dependence of $a_2$.
}
\begin{ruledtabular}
\begin{tabular}{c@{\extracolsep{0.5cm}}ccccc}
$\rho_S$ (g~cm$^{-3}$) & $\Phi$ & $\xi_{\|}$ (\AA) & 
$a_2 / \sigma_1$ ($\times 10^{-6}$\AA$^{-3}$) & $T_{IA}^*$ (K) & $x$\\
\hline
 0.0 & 1 &- & - & 324.49 & - \\
 0.025 & 0.99 & $4786 \pm 85$ & $0.071 \pm 0.003$ & $323.6 \pm 0.4$ & $0.14 \pm 0.01$ \\
 0.051 & 0.98 & $1927 \pm 24$ & $1.13 \pm 0.05$ & $323.8 \pm 0.8$ & $0.19 \pm 0.03$ \\
 0.105 & 0.95 & $1396 \pm 9$ & $1.74 \pm 0.04$ & $323.2 \pm 0.2$ & $0.39 \pm 0.15$ \\
 0.226 & 0.91 & $806 \pm 8$ & $10.8 \pm 0.9$ & $323.8 \pm 0.1$ & $0.22 \pm 0.02$ \\
 0.282 & 0.89 & $804 \pm 29$ & $2.35 \pm 0.30$ & $323.0 \pm 0.1$ & $0.32 \pm 0.06$ \\
 0.355 & 0.86 & $641 \pm 6$ & $12.2 \pm 0.9$ & $323.6 \pm 0.3$ & $0.51 \pm 0.20$ \\
 0.489 & 0.82 & $445 \pm 9$ & $40.3 \pm 20.4$ & $323.5 \pm 0.3$ & $0.39 \pm 0.10$ \\
 0.647 & 0.77 & $543 \pm 24$ & $7.62 \pm 2.99$ & $322.7 \pm 0.1$ & $0.41 \pm 0.13$ \\
\end{tabular}
\end{ruledtabular}
\end{table*}

The variation of the two components of the line shape, with temperature and $\rho_S$ is shown in
Fig.~\ref{4pictures}. Panels Fig.~\ref{4pictures}(a \& b) are for high temperature and
Fig.~\ref{4pictures}(c \& d) are for low temperature. While Fig.~\ref{4pictures}(a \& c) are for low
$\rho_S$ and Fig.~\ref{4pictures}(b \& d) are for high $\rho_S$. The second term in
Eq.(\ref{two_terms}) dominates the scattering for all temperatures studied (dashed line). The full
model Eq.(\ref{two_terms}) using Eq.(\ref{N_to_SmA}) for $S^{T}(\mathbf{q})$ 
is the solid line. The peak width varies little
with temperature but is seen to be broader than the instrumental resolution (Fig.~\ref{4pictures}).

\begin{figure}[b]
\includegraphics[scale=0.5]{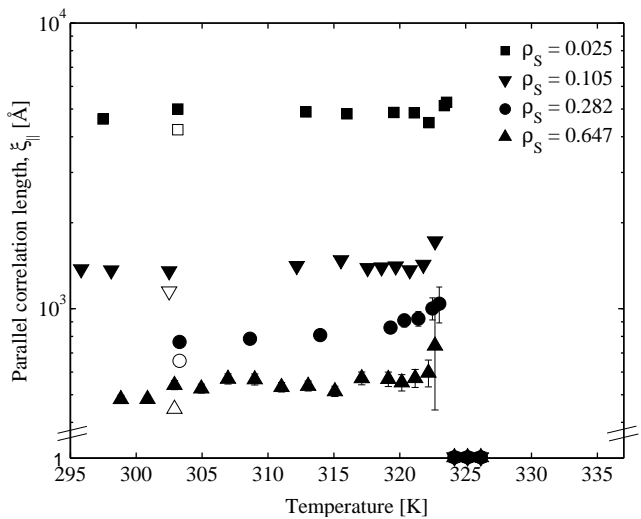}
\caption{\label{corr_T} The parallel correlation length, $\xi_{\|}$, for smectic
order as a function of the temperature for four 10CB-aerosil samples. These values denoted
by solid symbols are determined
using Eq.(\ref{N_to_SmA}) for $S^T(\mathbf{q})$ in 
Eq.(\ref{two_terms}) while the four open symbols correspond to
substituting for $S^T(\mathbf{q})$ given by Eq.(\ref{I_to_SmA}) into Eq.(\ref{two_terms}).}
\end{figure}

\begin{figure}[b]
\includegraphics[scale=0.5]{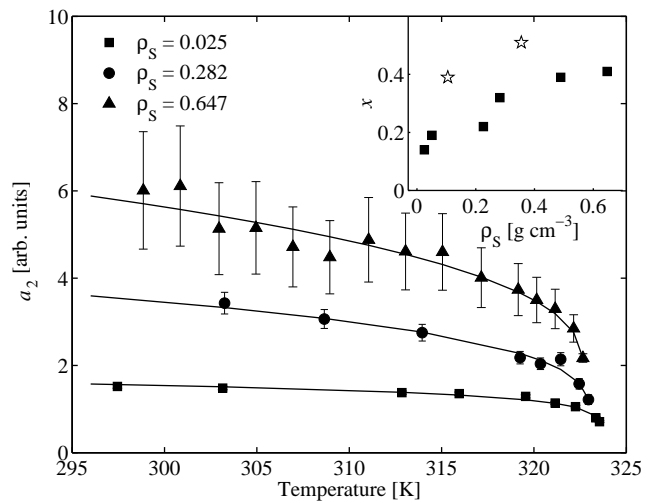}
\caption{\label{amp_T} Integrated intensity of the random-field term $S^{RF}(\mathbf{q})$ versus
temperature. The solid lines are the results of fits to a $(T_{IA}^* - T)^{x}$ 
power law form. The $x$ values are
shown in the inset and in Table~\ref{tab:Tna} along with the pseudotransition temperatures. The
stars in the inset indicate the most uncertain $x$ values.}
\end{figure}

Figures~\ref{corr_T} and \ref{amp_T} show parameters determined by the fit procedure versus temperature
while Figs.~\ref{corr_rho} and \ref{amp_rho} show parameters versus $\rho_S$. In the latter case, the
results are compared to those from other liquid crystal-aerosil gels at temperatures 
below the N-SmA transition.

The behavior of the parallel correlation length, $\xi_{\|}$, as a function of 
temperature is shown for a
selection of samples in Fig.~\ref{corr_T} (solid symbols). The fact that these are all finite 
shows that the quasi long-range SmA
state has been destroyed by the gel. The pseudotransition to short-range smectic order remains
sharp and the correlation length jumps directly to its low temperature value with no
pretransitional fluctuations. This behavior strongly suggests that the transition remains first
order in the gels. As the gel density is increased the correlation length systematically
decreases. The open symbols in Fig.~\ref{corr_T} show the values for $\xi_{\|}$ determined using
Eq.(\ref{two_terms}) with Eq.(\ref{I_to_SmA}) for $S^{T}(\mathbf{q})$. 
The systematic decrease in $\xi_{\|}$ results from the narrower
transverse profile of Eq.(\ref{I_to_SmA}) for a particular $\xi_{\bot}$ value. Since the shifts are
approximately equal on a logarithmic scale this result does not modify any of the power law 
relationships discussed in Sec.~\ref{sec:dis}.

The corresponding temperature dependence of the amplitude of the random-field term, $a_2$, is shown in
Fig.~\ref{amp_T} for three samples. 
The temperature dependence is consistent with the random field fluctuations
increasingly dominating the scattering at low temperatures. In Sec.~\ref{sec:dis} we will discuss
the difference between the behaviors of $a_2(T)$ and $\xi_{\|}(T)$.
In contrast to the correlation length behavior, the
growth of this amplitude parameter becomes increasingly more gradual as $\rho_S$ increases. 
As with N-SmA transitions in aerosil gels~\cite{Park},
$a_2(T)$ can be compared to
an order parameter squared for the pseudotransition. The lines in Fig.~\ref{amp_T} are the results
of fits with the form $a_2 \sim (T_{IA}^* - T)^{x}$. We emphasize that this form, which is normally
used at a second-order transition with $x = 2\beta$, is used here to provide a strictly
phenomenological description of the data. The inset in Fig.~\ref{amp_T} shows the $x$
values, and both they and the resulting pseudotransition temperatures, $T_{IA}^*$, are presented in
Table~\ref{tab:Tna}. Although there is substantial scatter, the $x$ values indicate that $a_2$
is growing more gradually with increasing $\rho_S$.

Figure~\ref{corr_rho} shows a log-log plot of the variation of the smectic domain size with the
gel density, $\rho_S$. The smectic domain size is quantified by an isotropic correlation length
$\overline{\xi} = (\xi_{\|}\xi_{\bot}^2)^{1/3}$ and the low-temperature average of this quantity
is shown. The line through the 10CB-aerosil data corresponds to $\overline{\xi} \sim \rho_S^{-1/2}$. 
Also plotted in
Fig.~\ref{corr_rho} are $\overline{\xi}$ results for 8CB-aerosil~\cite{Aerosil_X} and 
8OCB-aerosil~\cite{Clegg1} samples, for which
$\overline{\xi} \sim \rho_S^{-1}$. This contrast will be discussed in Sec.~\ref{sec:dis}.

Figure~\ref{amp_rho} is a log-log plot of $a_2 / \sigma_1$ versus the disorder strength as represented
by $\rho_S$. This ratio
gives the strength of the random-field contribution to the scattering and is independent of
x-ray intensity normalization.
Equivalent values are plotted for 8CB-aerosil~\cite{Aerosil_X} 
and 8OCB-aerosil~\cite{Clegg1} samples. While the scatter is rather large,
the results are consistent with power law behavior with $a_2 / \sigma_1 \sim \rho_S^3$ for N-SmA
and $a_2 / \sigma_1 \sim \rho_S^{3/2}$ for the I-SmA. The low temperature values of $a_2 /
\sigma_1$ and $\xi_{\|}$ are included in Table~\ref{tab:Tna}. These are the mean values below $315$
K.

\section{Discussion and Conclusions}
\label{sec:dis}

Our results show that an aerosil gel reduces the SmA state to short-range order with random-field
characteristics.
At first sight, the temperature dependencies in Fig.~\ref{corr_T} and \ref{amp_T} are a little
surprising. The behavior of the correlation length, $\xi_{\|}$, suggests a first-order transition 
for all $\rho_S$ while the
random-field amplitude, $a_2$, suggests an increasingly gradual transition with increasing $\rho_S$.
For pure 10CB the pretransitional fluctuations are very weak and short range, Fig.~\ref{Bulk}. 
While the correlation
lengths are short the liquid crystal is unlikely to be strongly perturbed by the aerosil gel
environment. The perturbation will only become effective when the correlation length grows to 
length scales comparable with the
distance between gel strands. Hence the correlation length
grows quickly until the disorder becomes apparent. As $\rho_S$ increases and $\xi_{\|}$ decreases, 
the order parameter behavior
changes from rising rapidly at low disorder to developing gradually for high gel densities. The
$x$ values inset to Fig.~\ref{amp_T} are consistent with a gradual crossover from strongly first order
to tricritical with increasing aerosil density.
As shown in Table~\ref{tab:Tna}
the pseudotransition temperature, $T_{IA}^*$, varies little with $\rho_S$ for 10CB-aerosil. This
suggests that local variations in $\rho_S$ are unlikely to lead to two-phase coexistence for this
material. Hence the transition is not observed to be substantially smeared. The correlation
length, $\overline{\xi} = (\xi_{\|} \xi_{\bot}^2)^{1/3}$, is characterizing short-range SmA 
domains within a predominantly SmA sample.

\begin{figure}
\includegraphics[scale=0.5]{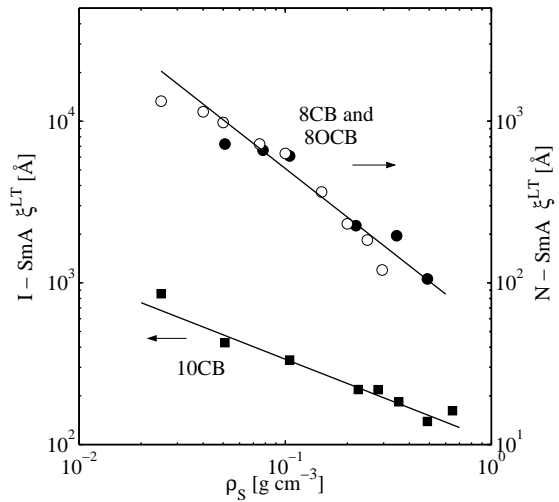}
\caption{\label{corr_rho} A comparison of the cube root of the correlation volume,
$\overline{\xi}$, taken at low temperature for 10CB-aerosil (solid squares),
8CB-aerosil~\cite{Aerosil_X} (open
circles) and 8OCB-aerosil~\cite{Clegg1} (solid circles). 
The last two bulk materials have N-SmA transitions. The solid lines represent simple
power law dependences, which differ for the two transitions.}
\end{figure}

\begin{figure}[b]
\includegraphics[scale=0.5]{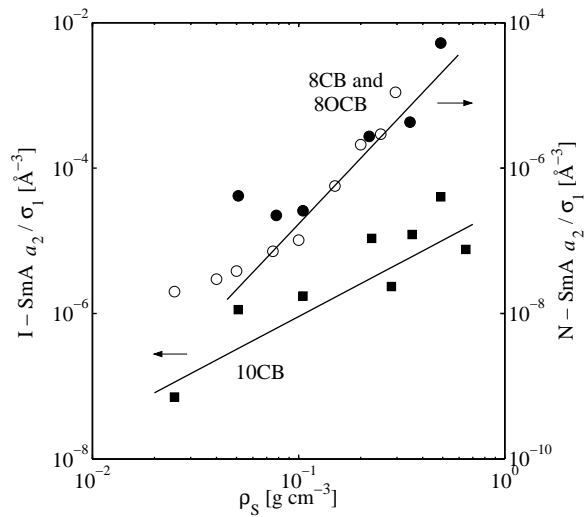}
\caption{\label{amp_rho} A comparison of the variation of $a_2 / \sigma_1$ versus $\rho_S$ for three
liquid crystal-aerosil systems at low temperatures. The
symbols are the same as Fig.~\ref{corr_rho}. The solid lines represent power law fits to the data
described in the text. The experimental scatter is obviously large.}
\end{figure}

Here we are interpreting our results as if two-phase coexistence over a range of temperatures were 
not an important issue. This
is because of the discontinuous behavior of $\xi_{\|}(T)$ at $T_{IA}^*$. By contrast, Bellini and
coworkers~\cite{Aerogel_IA}
interpreted their x-ray results on an 10CB-aerogel sample in terms of coexisting phases - indeed they
observed coexistence via optical microscopy. The 10CB-aerogel sample they studied had $\Phi \sim 0.79$
and $\overline{\xi} \sim 200$ \AA, roughly comparable with our highest $\rho_S$ aerosil sample. The
temperature dependence of the peak intensity and correlation lengths are similar. One important and
unusual feature of a fluctuation driven first order transition is that the disordered phase remains
metastable at all temperatures~\cite{Brazovskii, F_and_B}. 
It may well be the case that this leads to the observed coexistence.
This phenomena is also true of the bulk material and it is unlikely to have given rise to the change
in $a_2(T)$ with $\rho_S$ that we have presented.
The ratio of the volume of individual ordered domains to the volume of the 
complete SmA region is roughly 1:125000 and hence it is no
surprise that the remaining I phase plays little role in determining $\overline{\xi}$ and only a
very small role in the temperature dependence of the intensity.

The observed trend in $x$ (Fig.~\ref{amp_T}) together with previous results on other
cyanobiphenyls~\cite{Aerosil_Cp, Park,
Clegg1} suggests that the disorder introduces a systematic change in transition properties. For
8CB-aerosil~\cite{Aerosil_Cp, Park} and other N-SmA transitions the aerosils change the 
transition characteristics away from
an anisotropic critical point and toward 3D-XY behavior. The liquid crystal begins to behave as
though the N order parameter is becoming increasingly saturated. Here
10CB-aerosil appears to move toward tricritical properties ($x \sim 0.5$) as aerosil is added.
For the isotropic - nematic - smectic-A phase diagram this change is also equivalent to a shift
toward the regime where the N phase is stable. For 8CB-aerosil and 10CB-aerosil the disorder has
some effects which are equivalent to shortening the length of the molecule.

Up to this point the response of the I-SmA transition to the aerosil gel is very
similar to that of the N-SmA transition in aerosil gels~\cite{Park, Clegg1}. 
However, Figs.~\ref{corr_rho} and
\ref{amp_rho} show that there are systematic differences in the $\rho_S$ dependence for 
$\overline{\xi}$ and $a_2 / \sigma_1$ associated with these two transitions. In the case of N-SmA, it was argued
that $\rho_S$, the mass of aerosil divided by the volume of liquid crystal, corresponds to the
variance of the random field~\cite{Aerosil_X, Germano2}. 
The dependence of $\overline{\xi}$ and $a_2 / \sigma_1$ on the
random field variance $\Delta$ has been determined by Aharony and Pytte~\cite{A_and_P, Germano2}
\begin{equation}
\overline{\xi} \sim \Delta^{\frac{-1}{d_{\ell} - d}}
\label{corr_pow}
\end{equation}
\begin{equation}
a_2 / \sigma_1 \sim \Delta^{\frac{3}{d_{\ell} - d}}
\label{amp_pow}
\end{equation}

\noindent where $d_{\ell}$ is the lower marginal dimensionality in the presence of random fields. 
For transitions which break a continuous symmetry, random fields are found to shift the lower
marginal dimensionality to two higher dimensions. For
experimentally achievable resolutions the N-SmA transition has 3D-XY characteristics, hence $d_{\ell} =
4$. This is in full agreement with results for 8CB-aerosil~\cite{Aerosil_X} and 
8OCB-aerosil~\cite{Clegg1} as shown in
Figs.~\ref{corr_rho} and \ref{amp_rho}. These log-log plots strongly suggest that $d_{\ell} = 5$ for
10CB-aerosil. Two-phase coexistence on the scale of the individual ordered domains 
is predicted to give rise to more abrupt domain walls than
occur within an $n \geq 2$ single-phase sample~\cite{I_and_W}. 
By the Imry-Ma argument~\cite{I_and_M} this would tend to reduce
$d_{\ell}$ below 4. This is the opposite trend to our observations.

The observed $d_{\ell} = 5$ appears consistent with the absence of the orientationally ordered N phase. The
formation of SmA order in 10CB gives rise to nematic order and hence the director is constrained
to lie along the layer normal. Director fluctuations no longer influence the range of smectic
correlations. As a result the line shape has the form anticipated for the
Landau-Peierls instability Eq.(\ref{I_to_SmA}) albeit with finite-range order.
This is consistent with the use of Eq.(\ref{I_to_SmA}) as the thermal structure
factor for Eq.(\ref{two_terms}) and its square as the random-field term. The lower marginal
dimensionality is general for a transition from a uniform phase to one with translational order in
a single direction.
Golubovi\'{c} and Kuli\'{c}~\cite{G_and_K} have
made a theoretical analysis of a similar random-field problem. They studied a system with a
continuous set of energy minima in q-space, having the form of a ring as opposed to the sphere for
the I-SmA transition. Following Imry and Ma~\cite{I_and_M} they determined the dimensionalities for which
the correlation function for transverse fluctuations of the order parameter would diverge. The lower
marginal dimensionality depends on how the inverse susceptibility varies with the wave vector. For the XY
model the inverse susceptibility varies quadratically with the wave vector components~\cite{I_and_M}. 
For a ring of energy
minima the inverse susceptibility varies quartically with one of the wave vector components and
quadratically with the others~\cite{G_and_K}. 
For a sphere of energy minima the inverse susceptibility varies quartically
with two of the components and quadratically with the others. For $d \leq d_{\ell}$ the correlation function
in the presence of quenched random fields 
diverges and in these cases $d_{\ell} = 4, 4.5, 5$ respectively~\cite{I_and_M, G_and_K}.

The series of studies on 8CB-aerosil~\cite{Aerosil_Cp, Park, Aerosil_X, Germano2}, 
8OCB-aerosil~\cite{Clegg1} and 10CB-aerosil samples provide a detailed picture of
the effect of quenched random fields on the formation of the SmA phase. The first two examples
relate to the case where an orientationally ordered (N) phase intervenes between the I and the SmA
while the results presented here relate to the direct transition from I to SmA. The interaction
between nematic fluctuations away from the smectic layer normal
and the SmA order results in transverse smectic correlations which die
out with $q_{\bot}^{-2}$. Our results suggest that, at the I-SmA transition, the molecules are
constrained to be perpendicular to the layers
and hence the SmA correlations die out with $q_{\bot}^{-4}$. There is a
concomitant change in the lower marginal dimensionality.
We have presented high resolution x-ray diffraction results for 10CB-aerosil samples with a range
of gel densities. We have shown that an aerosil gel perturbs the I-SmA transition as though it
were a random field pinning the phase of the density wave. The transition remains first order for
all gel densities studied while the line shape and correlation length evolve systematically. The
line shape analysis reveals reasonable agreement with a Landau-Peierls system with quenched
random fields. It would be of very great interest to study the effect of aerosils at the tricritical
point close to the
boundary between N-SmA and I-SmA regimes.

\section{Acknowledgments}
Funding in Toronto was provided by the Natural Science and Engineering Research Council. The
National Synchrotron Light Source, Brookhaven National Laboratory, is supported by the US
Department of Energy under Contract No. DE-AC02-98CH10886.


\begin{thebibliography}{100}
\bibitem{Peierls}
R.E.~Peierls, Helv. Phys. Acta Suppl. \textbf{7}, 81 (1934).

\bibitem{L_and_L}
L.D.~Landau and E.M.~Liftshitz, \textit{Statistical Physics}, 2nd ed. (Pergamon Press, New York,
1980).

\bibitem{G_and_N}
C.W.~Garland and G.~Nounesis, Phys. Rev. E \textbf{49}, 2964 (1994) and references cited therein.

\bibitem{N_and_T}
D.R.~Nelson and J.~Toner, Phys. Rev. B \textbf{24}, 363 (1981) and references cited therein.

\bibitem{Aerogel_Cp}
L.~Wu, B.~Zhou, C.W.~Garland, T.~Bellini, and D.W.~Schaefer, Phys. Rev. E \textbf{51}, 2157 (1995).

\bibitem{Aerosil_Cp}
G.S.~Iannacchione, C.W.~Garland, J.T.~Mang, and T.P.~Rieker, Phys. Rev. E \textbf{58}, 5966 (1998).

\bibitem{Aerogel_DNMR}
H.~Zeng, B.~Zalar, G.S.~Iannacchione, and D.~Finotello, Phys. Rev. E \textbf{60}, 5607 (1999).

\bibitem{Aerosil_DNMR}
T.~Jin and D.~Finotello, Phys. Rev. Lett. \textbf{86}, 818 (2001).

\bibitem{Aerogel_X}
T.~Bellini, L.~Radzihovsky, J.~Toner, and N.A.~Clark, Science \textbf{294}, 1074 (2001).

\bibitem{Aerosil_X}
R.L.~Leheny, S.~Park, R.J.~Birgeneau, J.-L.~Gallani, C.W.~Garland, and
G.S.~Iannacchione, Phys. Rev. E \textbf{67}, 011708 (2003).

\bibitem{Aerogel_IA}
T.~Bellini, N.A.~Clark, and D.R.~Link, J. Phys.: Condens. Matter \textbf{15}, S175 (2003).

\bibitem{dG_and_P}
P.G.~de~Gennes and J.~Prost, \textit{The Physics of Liquid Crystals}, 2nd ed. (Oxford University
Press, New York, 1993).

\bibitem{Brazovskii}
S.A.~Brazovskii, Sov. Phys. JETP \textbf{41}, 85 (1975).

\bibitem{Anisimov}
M.A.~Anisimov, \textit{Critical Phenomena in Liquids and Liquid Crystals} (Gordon and Breach,
Philadelphia, 1991).

\bibitem{Marynissen}
H.~Marynissen, J.~Thoen, and W.~Van~Dael, Mol. Cryst. Liq. Cryst., \textbf{97}, 149 (1983).

\bibitem{Thoen_p}
The I-SmA first-order latent heat for 10CB is $2830 \pm 50$ J mol$^{-1}$ and the integrated area of the
pretransitional $C_p$ wings above and below T$_{IA}$ is only $\sim 275 \pm 100$ J mol$^{-1}$. 
J.~Thoen (private communication).

\bibitem{Gohin}
A.~Gohin, C.~Destrade, H.~Gasparoux, and J.~Prost, J. Phys. (France), \textbf{44}, 427 (1983).

\bibitem{Thoen}
J.~Thoen, H.~Marynissen, and W.~Van~Dael, Phys. Rev. Lett. \textbf{52}, 204 (1984).

\bibitem{Germano2}
G.S.~Iannacchione, S.~Park, C.W.~Garland, R.J.~Birgeneau, and R.L.~Leheny, Phys. Rev. E
\textbf{67}, 011709 (2003).

\bibitem{Clegg1}
P.S.~Clegg, C.~Stock, R.J.~Birgeneau, C.W.~Garland, A.~Roshi, and G.S.~Iannacchione,
Phys. Rev. E \textbf{67}, 021703 (2003).

\bibitem{I_and_M}
Y.~Imry and S.-K.~Ma, Phys. Rev. Lett. \textbf{35}, 1399 (1975).

\bibitem{I_and_W}
Y.~Imry and M.~Wortis, Phys. Rev. B \textbf{19}, 3580 (1979).

\bibitem{Ocko}
B.M.~Ocko, R.J.~Birgeneau, J.D.~Litster, and M.E.~Neubert, Phys. Rev. Lett. \textbf{52}, 208
(1984).

\bibitem{Als-Nielsen}
J.~Als-Nielsen, J.D.~Litster, R.J.~Birgeneau, M.~Kaplan, C.R.~Safinya, A.~Lindegaard-Andersen, and
S.~Mathiesen, Phys. Rev. B \textbf{22}, 312 (1980).

\bibitem{Litster}
J.D.~Litster, R.J.~Birgeneau, M.~Kaplan, C.R.~Safinya, and J.~Als-Nielsen, in \textit{Ordering in
Strongly Fluctuating Condensed Matter Systems}, edited by T.~Riste (NATO ASI, Volume 50, 1979).

\bibitem{Caille}
A.~Caill\'{e}, C. R. Acad. Sci. Ser. B \textbf{274}, 891 (1972).

\bibitem{C_and_S}
H.J.~Coles and C.~Strazielle, Mol. Cryst. Liq. Cryst. \textbf{49}, 259 (1979).

\bibitem{Degussa}
Degussa Corp., Silica Devision, 65 Challenger Road, Ridgefield Park, NY 07660.

\bibitem{Park}
S.~Park, R.L.~Leheny, R.J.~Birgeneau, J.-L.Gallani, C.W.~Garland, and G.S.~Iannacchione, Phys. Rev.
E \textbf{65}, 050703(R) (2002).

\bibitem{BDH}
BDH, Product Information (1986).

\bibitem{Clegg2}
P.S.~Clegg, R.J.~Birgeneau, S.~Park, C.W.~Garland, G.S.~Iannacchione, R.L.~Leheny, and M.E.~Neubert,
Phys. Rev. E \textbf{68}, 031706 (2003).

\bibitem{Leheny_p}
Recent studies by Leheny and coworkers show that, for anisotropic 8CB-aerosil and 
$\overline{8}S5$-aerosil gels, the relationships
$\xi_{\bot}(\xi_{\|})$ and $c(\xi_{\|})$ change compared to those for the pure liquid crystal 
materials.
The same type of variation, if it were to occur here, would give rise to very subtle
changes in parameter values.

\bibitem{Ocko_Thesis}
B.M.~Ocko, Ph.D. thesis, Massachusetts Institute of Technology, 1984 (unpublished).

\bibitem{F_and_B}
G.H.~Fredrickson and K.~Binder, J. Chem. Phys. \textbf{91}, 7265 (1989).

\bibitem{A_and_P}
A.~Aharony and E.~Pytte, Phys. Rev. B, \textbf{27}, 5872 (1983).

\bibitem{G_and_K}
L.~Golubovi\'{c} and M.~Kuli\'{c}, Phys. Rev. B \textbf{37}, 7582 (1988).

\end{thebibliography}
\end{document}